# Eight models for coherence of radiation from incoherent sources and coherence of sunlight

MIKHAIL CHARNOTSKII
Mikhail.Charnotskii@gmail.com

**Abstract** Eight models for the coherence of the quasi-monochromatic light from spherical incoherent sources are constructed by placing incoherent monopole and dipole sources on the surface of a sphere, inside a ball and on a plane circular disk. All models allow relatively simple numerical calculations of coherence functions for arbitrary source sizes and positions of observation points. We show that the far-field regime for transverse coherence is formed at the distances larger than the source size, regardless of the wavelength. Wide angle far field coherences have simple analytical representations in terms of the spherical and cylindrical Bessel functions. For very large sources the main and few first lobes of coherence function is well represented by paraxial approximation, which reduces the number of models to three. Based on the solar images corresponding to these three models the ball model emerges as the most appropriate. Ball model can be modified to match the limb darkening observations.

## 1. Introduction

Coherence of light from extended incoherent source was discussed in the classical text [1]. Born and Wolf used the van Cittert-Zernike theorem (VZT) to derive an analytical expression for the degree of coherence of radiation from incoherent monopole sources uniformly distributed on a disk, which we will replicate in Section 4, Eq. (58). Based on this model they estimated the size of coherence area at 88% coherence level for the Sun light observed on Earth as approximately $34\lambda$, where $\lambda$ is radiation wavelength.

Several recent papers discussed coherence properties of light field created by incoherent spherical source, in application to the sunlight.

In [2] boundary condition in the form of delta-correlated field at the spherical surface was used. The coherence function (cross-spectral density) outside the sphere was represented in the form of infinite series in terms of spherical Hankel functions and Legendre polynomials. Authors argued that series can be truncated at about $ka$ terms, where $k = 2\pi/\lambda$ is wavenumber, and $a$ is the sphere radius. Based on numerical calculations for $ka = 100$, authors concluded that the far-field for the transverse Degree of Coherence (DOC) is formed at a distance of several $\lambda$ from the surface, and the shape of the far-field DOC is similar to the monopoles on the disc model, [1] and Eq. (58) here. Authors also stated that the series technique is inadequate for the sunlight coherence calculations due to the very large number of terms needed.

In [3] the same spherical-harmonics based approach was used for the partially coherent fields at a sphere, as was proposed in [4]. Additional, ad hoc, approximation replaced the infinite series summation by integration. For the delta-correlated boundary condition on DOC this led to the far-field paraxial DOC, Eq. (29), corresponding here to the incoherent monopole and isotropic dipole sources in the ball, and normal dipole sources on a sphere. The case of the black body boundary condition, resulted in the far-field paraxial DOC, Eq. (59), corresponding to the incoherent sources at a disk. Once again, authors limited their examples by the $ka = 100$ case and did not present results for the sunlight case where $ka \sim 10^{16}$.

In [5, 6] the VZT-based DOC calculations were extended to the broadband radiation by spectral integration of the far-field flat disk quasi-monochromatic DOC, Eq. (59) weighted by solar spectrum. In [7] the VZT calculations of the solar DOC included the limb darkening [8], and showed measurable differences vs the uniform disk model. The authors also accounted for the scattered sunlight by adding a diffuse component with black-body DOC. The interferometric measurements of the sunlight DOC in [7, 9] are in a reasonable agreement with the VZT estimates.

In [10] the far-field DOC of electromagnetic field created by quasi-homogeneous isotropic random currents was considered. For the Gaussian source it was found that the coherence angle in the far-field is inverse proportional to $ka$ and DOC has approximately Gaussian shape. In [11] the spherical harmonics expansion was used to calculate the coherence of the electromagnetic field created by delta-correlate source currents at the sphere. Electromagnetic DOC for a wide range of source sizes, $0.1 \leq ka \leq 500$ was considered based on numerical summation of the harmonic series. Again, it was observed that it is necessary to sum about $ka$ number of terms to calculate the far field DOC. It was also noted that the far-field DOC is a universal function of a certain parameter depending on the angular separation and $ka$. Also, while not explicitly stated, the author's comments suggests that the far-field is formed at the distances of several wave lengths from the surface.

In the recent paper [12], small-angle approximation, which replaced the Legendre polynomials by Bessel functions, was used in the spherical harmonics series of [2]. Authors state that their approximation is uniform in series terms numbers, and, similar to [3], replace the infinite summation by integration to arrive at the far-field paraxial DOC for the incoherent disk, Eq. (59). Without sufficient explanations, the far-field behavior of the DOC, quoted to form at distance of several $\lambda$ from the source, [2], is related to the small angle subtended by the observation points separation. As we show in section 5, far field for DOC is formed at distances $R > a$ both for the wide angle and paraxial cases.

The paper is organized as follows. In Section 2 we discuss coherence of the fields created by incoherent monopole and dipole sources at the spherical surface. Section 3 considers incoherent monopole and dipole sources distributed in a spherical ball. Section 4 considers incoherent monopole and dipole sources distributed at a plane disc. Section 5 discusses the far field conditions for incoherent sources. In Section 6 we discuss DOC models in the paraxial approximation, present calculations for coherence of solar light, and analyze Sun images corresponding to different models in relation to the limb darkening effect.

## 2. Spherical sources

Complex amplitude of scalar monochromatic field $U(\mathbf{R})$ created by monopole sources distribution $S(\mathbf{R})$ is described by Helmholtz equation

$$\Delta U(\mathbf{R}) + k^2 U(\mathbf{R}) = S(\mathbf{R}) \qquad (1)$$

Free-space solution of Eq. (1) is

$$U(\mathbf{R}) = \iiint d^3 R_s S(\mathbf{R}_s) G(\mathbf{R} - \mathbf{R}_s), \qquad (2)$$

where free-space Green's function

$$G(\mathbf{R}) = \frac{-1}{4\pi R} \exp(ikR), \qquad (3)$$

is solution for a monopole source at coordinate origin

$$\Delta G(\mathbf{R}) + k^2 G(\mathbf{R}) = \delta(\mathbf{R}), \qquad (4)$$

and $k = 2\pi/\lambda$ is the wavenumber.

Field of a unit dipole with axis direction $\hat{\mathbf{p}}$, corresponds to $S(\mathbf{R}) = \hat{\mathbf{p}} \cdot \nabla \delta(\mathbf{R})$, and is

$$U(\mathbf{R}) = -\hat{\mathbf{p}} \cdot \nabla G(\mathbf{R}) = -\hat{\mathbf{p}} \cdot \mathbf{R} \frac{(1-ikR)}{4\pi R^3} \exp(ikR) \tag{5}$$

In one special case the free-space Green's function allows to solve a boundary-value problem for Helmholtz equation. Specifically, if field is known at the plane $z = 0$ of coordinate system $\mathbf{R} = (\mathbf{r}, z)$, and all the sources are located in the half-space $z < 0$, Green's integral identity leads to representation of the field in the half-space $z > 0$, in terms of the boundary values of the field $U(\mathbf{r}_0, 0)$ at plane $z = 0$

$$U(\mathbf{r}, z) = 2 \iint d^2 r_0 U(\mathbf{r}_0, 0) \frac{\partial G(\mathbf{r} - \mathbf{r}_0, z)}{\partial z}, \quad z > 0. \tag{6}$$

In this section we analyze coherence of the field created by incoherent sources distributed on the surface of a sphere with radius $a$.

### 2.1 Monopole sources

Here we consider monopole sources located at the surface of a sphere with radius $a$. Field at the point $\mathbf{R}$ outside the sphere can be presented as

$$U(\mathbf{R}) = \frac{-1}{4\pi} \iint_{R_S=a} d^2 R_S \frac{m(\mathbf{R}_S)}{|\mathbf{R} - \mathbf{R}_S|} \exp(ik|\mathbf{R} - \mathbf{R}_S|), \tag{7}$$

where $m(\mathbf{R}_S)$ is complex amplitude of the source at the point $\mathbf{R}_S$

Coherence function of this field is

$$W(\mathbf{R}_1, \mathbf{R}_2) = \langle U(\mathbf{R}_1) U(\mathbf{R}_2) \rangle$$
$$= \frac{1}{16\pi^2} \iint_{R_S=a} d^2 R_S \iint_{R'_S=a} d^2 R'_S \frac{\langle m(\mathbf{R}_S) m^*(\mathbf{R}'_S) \rangle}{|\mathbf{R}_1 - \mathbf{R}_S||\mathbf{R}_2 - \mathbf{R}'_S|} \exp(ik|\mathbf{R}_1 - \mathbf{R}_S| - ik|\mathbf{R}_2 - \mathbf{R}'_S|), \tag{8}$$

For a simple case of incoherent sources distribution

$$\langle m(\mathbf{R}_S) m^*(\mathbf{R}'_S) \rangle = M(\mathbf{R}_S) \delta_{SPH}(\mathbf{R}_S - \mathbf{R}'_S), \tag{9}$$

where $M(\mathbf{R}_S)$ is the source surface power density and $\delta_{SPH}(\ )$ is delta-function on a sphere. In the spherical coordinate representation

$$\mathbf{R} = R(\cos\theta\cos\varphi\hat{\mathbf{x}} + \cos\theta\sin\varphi\hat{\mathbf{y}} + \sin\theta\hat{\mathbf{z}}),$$
$$\delta_{SPH}(\mathbf{R}_S - \mathbf{R}'_S) = \frac{\delta(\varphi - \varphi')\delta(\theta - \theta')}{a^2 \cos\theta}, \tag{10}$$

and the two-point coherence function of the field created by a uniform, $M(\mathbf{R}_S) = M$ sources distribution is presented as an integral over the unit sphere

$$W(\mathbf{R}_1, \mathbf{R}_2) = \frac{Ma^2}{16\pi^2} \int_{-\pi}^{\pi} d\varphi \int_{-\pi/2}^{\pi/2} \frac{d\theta \cos\theta}{|\mathbf{R}_1 - \mathbf{R}_S||\mathbf{R}_2 - \mathbf{R}_S|} \exp(ik|\mathbf{R}_1 - \mathbf{R}_S| - ik|\mathbf{R}_2 - \mathbf{R}_S|). \tag{11}$$

Wave irradiance is

$$I(R) = W(R, 0) = \frac{Ma^2}{8\pi} \int_{-\pi/2}^{\pi/2} \frac{d\theta \cos\theta}{a^2 + R^2 - 2aR\sin\theta} = \frac{Ma}{8\pi R} \ln\left(\frac{R+a}{R-a}\right). \tag{12}$$

Irradiance is quasi static, that is bears no dependence on the wave number, and diverges for $R \to +a$, which is a result of delta-correlated source model. For $R \gg a$

$$I(R \gg a) \approx \frac{Ma^2}{4\pi R^2}, \qquad (13)$$

matching the expected far-field $R^{-2}$ behavior. However the $I(R)R^2$ product is not constant in general. Power flux density is

$$\mathbf{J}(\mathbf{R}) \propto \mathrm{Im}\langle \nabla U(\mathbf{R}) U^*(\mathbf{R}) \rangle = \mathrm{Im}\, \nabla_1 W(\mathbf{R}_1, \mathbf{R}_2)_{\mathbf{R}_1 = \mathbf{R}_2 = \mathbf{R}} = \frac{kMa^2}{4\pi R^2} \hat{\mathbf{R}}, \qquad (14)$$

and total power flux through any surface surrounding $\mathbf{R}=0$ is constant.

In the case of transverse coherence function without a loss of generality we can set

$$\mathbf{R}_1 = R\left(\hat{\mathbf{x}} \sin\frac{\alpha}{2} + \hat{\mathbf{z}} \cos\frac{\alpha}{2}\right), \quad \mathbf{R}_2 = R\left(-\hat{\mathbf{x}} \sin\frac{\alpha}{2} + \hat{\mathbf{z}} \cos\frac{\alpha}{2}\right), \qquad (15)$$

where $\alpha$ is the angle between the vectors $\mathbf{R}_1$ and $\mathbf{R}_2$, and $W(\mathbf{R}_1,\mathbf{R}_2)=W(R,\alpha)$. Transverse DOC is presented as a two-fold integral

$$w(R,\alpha) = \frac{W(R,R,\alpha)}{I(R)} = \frac{aR}{2\pi \ln\left(\frac{R+a}{R-a}\right)} \int_{-\pi}^{\pi} d\varphi \int_{-\pi/2}^{\pi/2} \frac{\cos\theta\, d\theta\, \cos\{k[r(\alpha)-r(-\alpha)]\}}{r(\alpha)r(-\alpha)}, \qquad (16)$$

where we used a shorthand notation

$$r(\alpha) \equiv \left[a^2 + R^2 - 2aR\left(\cos\varphi \cos\theta \sin\frac{\alpha}{2} + \sin\theta \cos\frac{\alpha}{2}\right)\right]^{1/2}. \qquad (17)$$

Clearly, DOC is a function of two dimensionless parameters, say $ka$ and $R/a$, and angular separation $\alpha$. For finite values of parameters DOC can be readily calculated numerically using Eq. (16). Fig. 1 presents examples of DOC calculated by numeric integration of Eq. (16).

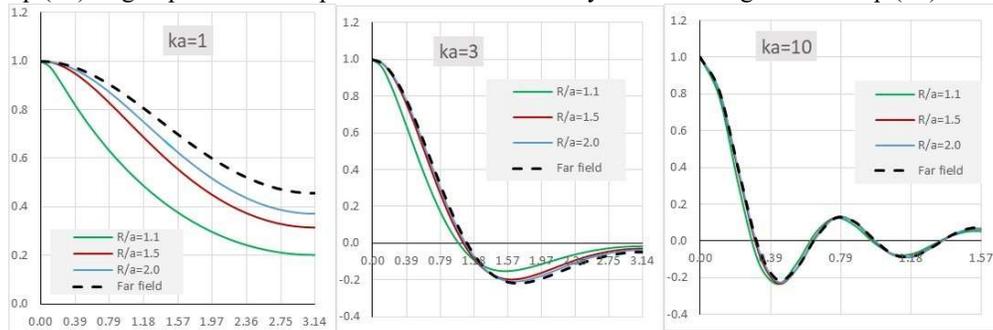

Fig. 1. Solid curves - numerically calculated DOCs for monopole sources at a sphere, Eq. (16), for three values of $ka$. Dashed curves – far-field result, Eq. (19).

Fig. 1, and additional calculations that are not displayed here, indicate that for $ka \leq 1$ field maintains positive coherence at any distances from the source and point separations. For $ka>1$ the width of the main lobe of coherence function (coherence angle) decreases in inverse proportion to $ka$. It is also notable that for $ka>3$ dependence of DOC as a function of separation

angle $\alpha$ on the distance from the source is quite weak. Coherence angle does not tend to zero when points approach the source, and present, incoherent source model, does not reproduce the incoherent field assumption of the previous calculations [2, 3].

Numerical integration of the double integral in Eq. (16) can become problematic for extreme values of $ka$ and $kR$. Meanwhile, in these cases corresponding asymptotes can be calculated analytically. For $R \gg a$ Eq. (16) can be simplified by using expansions

$$r(\alpha)r(-\alpha) = R^2 \left[1 + O\left(\frac{a}{R}\right)\right],$$
$$r(\alpha) - r(-\alpha) = -2a\cos\varphi\cos\theta\sin\frac{\alpha}{2}\left[1 + \frac{a}{R}\sin\theta\cos\frac{\alpha}{2} + O\left(\frac{a^2}{R^2}\right)\right] \quad (18)$$

and with the help of Eq. (6.567.13) of [14] Eq. (16) simplifies to

$$w_{FF}(\alpha) = \frac{1}{2}\int_{-\pi/2}^{\pi/2} \cos\theta \, d\theta \, J_0(\gamma\cos\theta) = j_0(\gamma) = \frac{\sin\gamma}{\gamma}, \quad \gamma = 2ka\sin\frac{\alpha}{2}. \quad (19)$$

Dashed curves in Fig. (1) show far-field results, Eq. (19). For $ka > 1$ $w(\alpha)$ maintains finite values only for $\alpha < 1$. In this case the "triple-sine" result of Eq. (19) can be simplified to a paraxial "sinc" form

$$w_{PA}(\alpha) \approx \frac{\sin(ka\,\alpha)}{ka\,\alpha}. \quad (20)$$

Fig. 2 shows the examples of the wide angle far-field DOC, Eq. (19) as solid curves and paraxial approximation, Eq. (20) for different values of parameter $ka$. As expected, "sinc" approximation is adequate for the practically important central part of DOC when $ka > 1$.

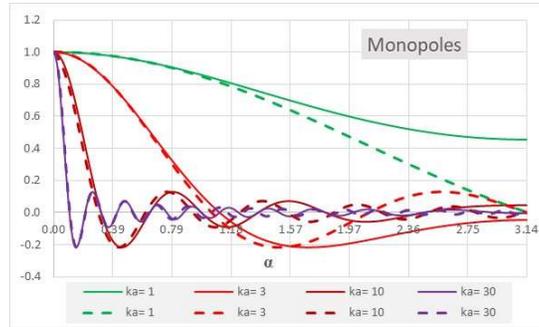

Fig. 2. Far-field DOC $w_{FF}(\alpha)$ for monopole sources for different values of $ka$. Solid curves, Eq. (19), dashed lines - paraxial approximation, Eq. (20).

As should be expected for a far-field solution, DOC depends on the angular measure of the point separation only. The crucial question remaining is the validity domain of the far-field DOC, Eq. (19). This issue will be discussed in Section 5.

### 2.2 Normal dipoles

Here we consider dipole sources located at the surface of a sphere with radius $a$. Dipole axis is assumed to be normal to the surface. Field at the point $\mathbf{R}$ outside the sphere can be presented as

$$U(\mathbf{R}) = \frac{1}{4\pi a}\iint_{R_S=a} d^2 R_S \frac{d_N(\mathbf{R}_S)(a^2 - \mathbf{R}_S \cdot \mathbf{R})}{|\mathbf{R} - \mathbf{R}_S|^3}\left[1 - ik|\mathbf{R} - \mathbf{R}_S|\right]\exp(ik|\mathbf{R} - \mathbf{R}_S|), \quad (21)$$

where $d_N(\mathbf{R}_S)$ is complex amplitude of dipole at the point $\mathbf{R}_S$.

Assuming uniform distribution of incoherent sources on the surface,

$$\langle d_N(\mathbf{R}_S) d_N^*(\mathbf{R'}_S) \rangle = D_N \delta_{SPH}(\mathbf{R}_S - \mathbf{R'}_S), \tag{22}$$

and using transformation similar to the monopole case we can present coherence function of the field as

$$W(\mathbf{R}_1, \mathbf{R}_2) = \frac{D_N a^2}{16\pi^2} \int_{-\pi}^{\pi} d\varphi \int_{-\pi/2}^{\pi/2} \frac{d\theta \cos\theta (\mathbf{R}_1 \cdot \mathbf{R}_S - a^2)(\mathbf{R}_2 \cdot \mathbf{R}_S - a^2)}{|\mathbf{R}_1 - \mathbf{R}_S|^3 |\mathbf{R}_2 - \mathbf{R}_S|^3}$$
$$(1 - ik|\mathbf{R}_1 - \mathbf{R}_S|)(1 + ik|\mathbf{R}_2 - \mathbf{R}_S|) \exp(ik|\mathbf{R}_1 - \mathbf{R}_S| - ik|\mathbf{R}_2 - \mathbf{R}_S|). \tag{23}$$

and calculate the irradiance

$$I(R) = W(\mathbf{R}, \mathbf{R}) = \frac{D_N a^2}{8\pi} \int_{-\pi/2}^{\pi/2} \frac{d\theta \cos\theta (a^2 - aR\sin\theta)^2}{(a^2 + R^2 - 2aR\sin\theta)^3} \left[1 + k^2(a^2 + R^2 - 2aR\sin\theta)\right]$$
$$= \frac{D_N}{32\pi} \left[ \frac{a}{R} \ln\left(\frac{R+a}{R-a}\right) - 2a^2 \frac{R^2 - 3a^2}{(R^2 - a^2)^2} \right] + \frac{D_N}{16\pi} k^2 a^2 \left[ 2 - \frac{(R^2 - a^2)}{aR} \ln\left(\frac{R+a}{R-a}\right) \right]. \tag{24}$$

For $R \gg a$ Eq. (24) reduces to

$$I(R \gg a) \approx \frac{D_N a^4}{12\pi R^4}(1 + k^2 R^2), \tag{25}$$

Unlike the monopole case, Eq. (13), irradiance is characterized by two components. First, quasi-static component, is frequency-independent and decreases as $1/R^4$. The second, "wave" term, is proportional to $(ka)^2$, and decreases much slower as $1/R^2$. Fig. 3 shows dependence of $I(R)(a/R)^2$ on parameters $ka$ and $r/a$. It confirms that the quasi-static regime for irradiance exists only at the distances of several wave lengths from the source. For a typical optical case the second term is dominant. The $I(R)R^2$ product is not constant in general, but the power flux density is

$$\mathbf{J}(\mathbf{R}) \propto \text{Im}\langle \nabla U(\mathbf{R}) U^*(\mathbf{R}) \rangle = \text{Im}\nabla_1 W(\mathbf{R}_1, \mathbf{R}_2)_{\mathbf{R}_1 = \mathbf{R}_2 = \mathbf{R}} = \frac{k^3 a^4 D_N}{12\pi R^2} \hat{\mathbf{R}}, \tag{26}$$

and total power flux through any surface enclosing the source is constant.

Transverse DOC for normal-oriented dipoles is

$$w(R,\alpha) = \frac{a^2 D_N}{16\pi^2 I(R)} \int_{-\pi}^{\pi} d\varphi \int_{-\pi/2}^{\pi/2} \frac{\cos\theta \, d\theta}{[r(\alpha)r(-\alpha)]^3} [1 - ikr(\alpha)][1 + ikr(-\alpha)] \exp\{ik[r(\alpha) - r(-\alpha)]\}$$
$$\left[ a^2 - ar\left(\cos\varphi\cos\theta\sin\frac{\alpha}{2} + \sin\theta\cos\frac{\alpha}{2}\right)\right]\left[ a^2 - ar\left(-\cos\varphi\cos\theta\sin\frac{\alpha}{2} + \sin\theta\cos\frac{\alpha}{2}\right)\right], \tag{27}$$

Figure 4 shows examples of DOC for several values of $ka$. In contrast to the monopole case, coherence vanishes for equatorially spaced points for $ka \propto 1$. However, the width of the main lobe of coherence function decreases in inverse proportion to $ka$ similar to the monopole case.

Eq. (27) can be simplified for $R > a$ by using expansions, Eq. (18) as follows

$$w_{FF}(\alpha) = \frac{3}{2}\cos^2\frac{\alpha}{2}\int_{-\pi/2}^{\pi/2}\cos\theta\sin^2\theta J_0(\gamma\cos\theta)d\theta - \frac{3}{4}\sin^2\frac{\alpha}{2}\int_{-\pi/2}^{\pi/2}\cos^3\theta[J_0(\gamma\cos\theta) - J_2(\gamma\cos\theta)]d\theta \quad (28)$$

$$= 3\cos^2\frac{\alpha}{2}\left[\frac{\sin\gamma}{\gamma^3} - \frac{\cos\gamma}{\gamma^2}\right] - 3\sin^2\frac{\alpha}{2}\left[\frac{\sin\gamma}{\gamma} - 2\frac{\sin\gamma}{\gamma^3} + 2\frac{\cos\gamma}{\gamma^2}\right], \quad \gamma = 2ka\sin\frac{\alpha}{2}$$

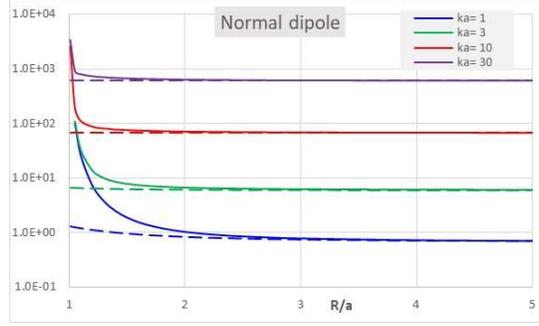

Fig. 3. Irradiance $I(R)$ from incoherent normal dipoles on a sphere for different values of $ka$.
Solid curves, Eq. (24), dashed lines approximation, Eq. (25).

Dashed curves in Fig. 4 show far-field results, Eq. (28). For $ka > 1$ $w(\alpha)$ maintains finite values only for $\alpha < 1$. In this case Eq. (28) equation can be simplified to

$$w_{PA}(\alpha) = 3\frac{j_1(ka\alpha)}{(ka\alpha)} = 3\left[\frac{\sin(ka\alpha)}{(ka\alpha)^3} - \frac{\cos(ka\alpha)}{(ka\alpha)^2}\right]. \quad (29)$$

Eq. (29) was derived in [3] based on approximate summation of the spherical harmonics series for DOC satisfying delta-correlated boundary conditions at the spherical surface. Fig. 5 shows the examples of the wide-angle far-field DOCs, Eq. (28) as solid curves and the paraxial approximation, Eq. (29), for different values of $ka$. As expected, Eq. (29) is adequate for the central part of DOC when $ka > 1$.

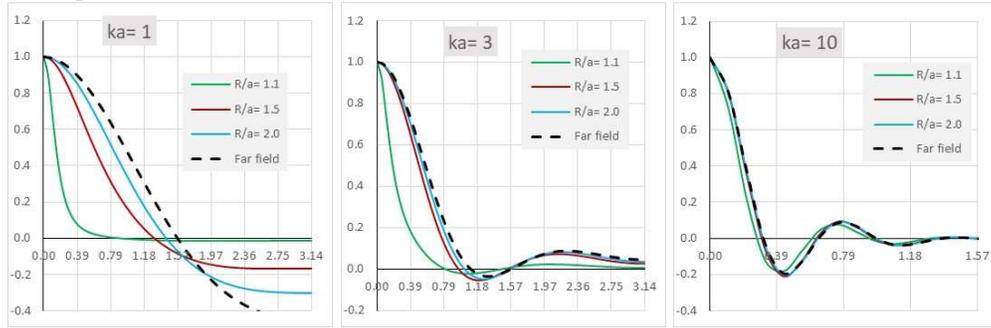

Fig. 4. Solid curves - numerically calculated DOC for incoherent normal dipoles on a sphere,
Eq. (28), for three values of $ka$. Dashed curves – far-field result, Eq. (29).

## 2.3 Isotropic dipoles

Here we consider randomly oriented dipole sources on a sphere. Field at the point $\mathbf{R}$ outside the sphere can be presented as

$$U(\mathbf{R}) = \frac{1}{4\pi}\iint_{R_S=a} d^2R_S \frac{d_I(\mathbf{R}_S)\hat{\mathbf{d}}_I(\mathbf{R}_S)\cdot(\mathbf{R}_S - \mathbf{R})}{|\mathbf{R} - \mathbf{R}_S|^3}[1 - ik|\mathbf{R} - \mathbf{R}_S|]\exp(ik|\mathbf{R} - \mathbf{R}_S|). \quad (30)$$

Here $d_I(\mathbf{R}_S)$ is random complex amplitude, and $\hat{\mathbf{d}}_I(\mathbf{R}_S)$ is a random orientation of dipole momentum at the point $\mathbf{R}_S$. Assuming uniform distribution of incoherent complex amplitudes on the surface, similar to Eq. (22), and independent isotropic distribution of orientations, we calculate coherence function as

$$W(\mathbf{R}_1,\mathbf{R}_2) = \frac{D_I a^4}{48\pi^2} \int_{-\pi}^{\pi} d\varphi \int_{-\pi/2}^{\pi/2} \frac{d\theta \cos\theta (\mathbf{R}_1-\mathbf{R}_S)\cdot(\mathbf{R}_2-\mathbf{R}_S)}{|\mathbf{R}_1-\mathbf{R}_S|^3 |\mathbf{R}_2-\mathbf{R}_S|^3}$$
$$(1-ik|\mathbf{R}_1-\mathbf{R}_S|)(1+ik|\mathbf{R}_2-\mathbf{R}_S|)\exp(ik|\mathbf{R}_1-\mathbf{R}_S|-ik|\mathbf{R}_2-\mathbf{R}_S|).$$ (31)

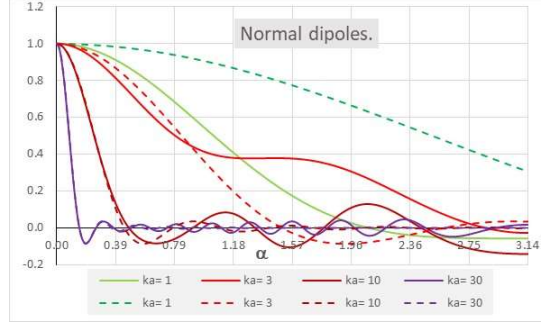

Fig. 5. Far-field DOC for incoherent normal dipoles on a sphere for different values of $ka$.
Solid curves, Eq. (28), dashed lines paraxial approximation, Eq. (29).

Irradiance can be calculated from Eq. (31) for $\mathbf{R}_1 = \mathbf{R}_2$ as

$$I(R) = W(\mathbf{R},\mathbf{R}) = \frac{D_I a^4}{24\pi} \int_{-\pi/2}^{\pi/2} d\theta \cos\theta \left[ \frac{1}{(a^2+R^2-2aR\sin\theta)^2} + \frac{k^2}{(a^2+R^2-2aR\sin\theta)} \right]$$
$$= \frac{D_I}{12\pi}\left[\frac{a^4}{(R^2-a^2)^2} + k^2 a^2 \frac{a}{2R}\ln\left(\frac{R+a}{R-a}\right)\right].$$ (32)

For $R \gg a$ Eq. (32) reduces to Eq. (25) for normal dipoles. Similar to the normal dipole case, Eq. (25), irradiance is characterized by two components: quasi-static and "wave" terms. Fig. 6 shows dependence of $I(R)(a/R)^2$ on $ka$ and $r/a$. The dependencies are very similar to the normal dipoles case.

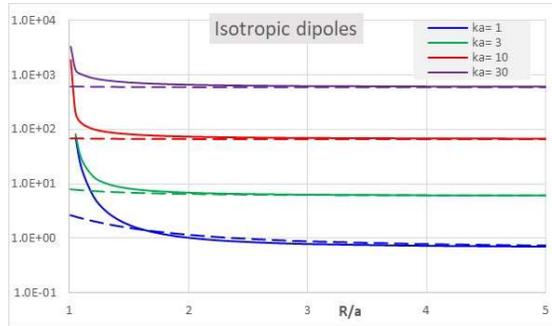

Fig. 6. Irradiance $I(R)$ from incoherent isotropic dipoles on a sphere for different values of $ka$.
Solid curves, Eq. (32), dashed lines approximation, Eq. (25).

Transverse DOC for normal dipoles is

$$w(R,\alpha) = \frac{a^2 D_I}{48\pi^2 I(R)} \int_{-\pi}^{\pi} d\varphi \int_{-\pi/2}^{\pi/2} \frac{\cos\theta\, d\theta}{[r(\alpha)r(-\alpha)]^3}[1-ikr(\alpha)][1+ikr(-\alpha)]\exp\{ik[r(\alpha)-r(-\alpha)]\}$$
$$\times \left[a^2 - 2aR\sin\theta\cos\frac{\alpha}{2} + R^2\cos\alpha\right], \quad (33)$$

Figure 7 shows examples of DOCs for several values of $ka$. Similar to the two previous cases, the width of the main lobe of coherence function decreases in inverse proportion to $ka$.

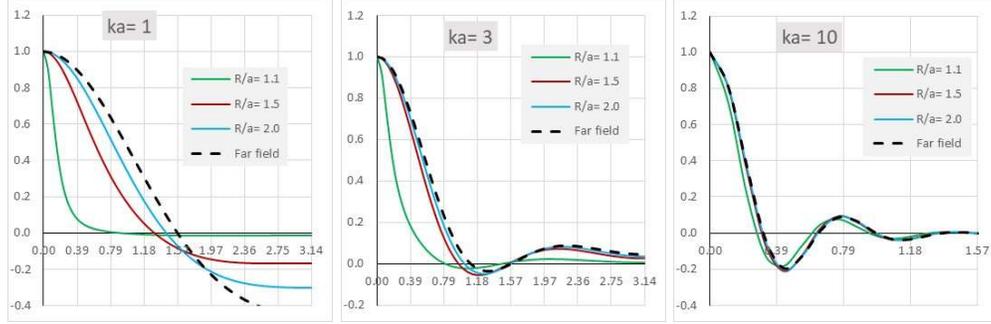

Fig. 7. Solid curves - numerically calculated DOCs for incoherent isotropic dipoles on a sphere, Eq. (33), for three values of $ka$. Dashed curves – far-field result, Eq. (34).

Eq. (33) can be simplified for $R > a$ by using expansions, Eq. (18) as follows

$$w_{FF}(\alpha) = \frac{1}{2}\cos\alpha \int_{-\pi/2}^{\pi/2} \cos\theta J_0(\gamma\cos\theta)d\theta = \cos\alpha\,\frac{\sin\gamma}{\gamma}, \quad \gamma = 2ka\sin\frac{\alpha}{2} \quad (34)$$

Dashed curves in Fig. 7 show far-field results, Eq. (34). For $ka > 1$ DOC maintains finite values only for $\alpha < 1$. In this case Eq. (34) equation can be simplified to the paraxial form given by Eq. (29). Fig. 8 shows the examples of the wide-angle far-field DOCs, Eq. (34) as solid curves and the paraxial approximation, Eq. (29), for different values of $ka$. As expected, Eq. (29) is adequate for the central part of DOC when $ka > 1$.

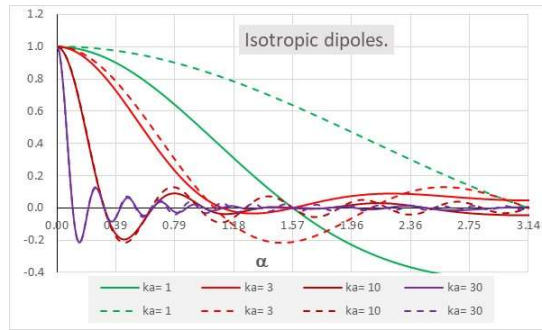

Fig. 8. Far-field DOC $w_{FF}(\alpha)$ for incoherent normal dipoles on a sphere for different values of $ka$. Solid curves, Eq. (34), dashed lines paraxial approximation, Eq. (29).

Fig. 9 compares the far-field DOC for the three model spherical sources. There are substantial difference between the models for moderate $ka$ values, but models become very similar with for large $ka$ values when $ka\alpha \propto 1$ with notable exception of the normal dipole model.

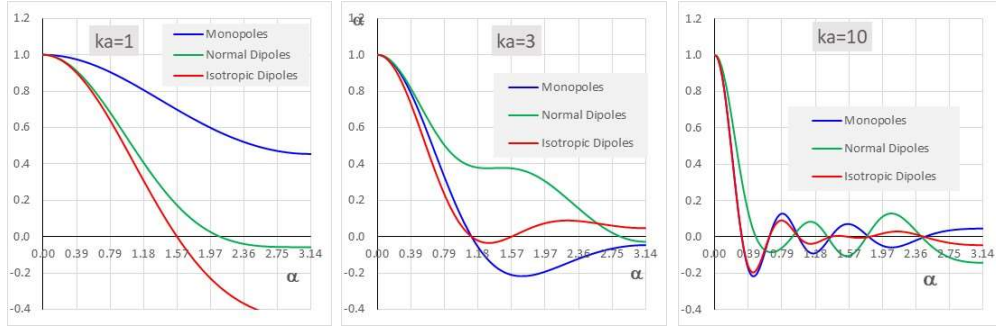

Fig. 9. Far-field DOC for three types of incoherent sources on a sphere, Eq. (19), Eq. (28) and Eq. (34) for different values of $ka$.

## 3. Ball source

In this section we analyze coherence of the field created by incoherent sources distributed inside the ball with radius $a$. Only monopole and randomly oriented dipoles are considered. Obvious generalization of the normal surface dipoles of the previous section would be radially oriented dipoles in a ball. However uniform dipole strength throughout the ball, would cause a singularity at the ball center, and it would be necessary to introduce radial dependence $D(P)$ with $D(0)=0$ to mitigate this singularity. Without having any physical models for $D(P)$ we decided to not include the radial dipoles here.

### 3.1 Monopole sources

For monopole sources filling the ball with radius $a$ field at the point $\mathbf{R}$ outside the sphere can be presented as

$$U(\mathbf{R}) = \frac{-1}{4\pi} \iiint_{P \leq a} d^3 P \frac{m(\mathbf{P})}{|\mathbf{R}-\mathbf{P}|} \exp(ik|\mathbf{R}-\mathbf{P}|), \tag{35}$$

where $m(\mathbf{P})$ is complex amplitude of the source at the point $\mathbf{P}$.

Similar to the previous section we consider a of uniform distribution of incoherent sources inside the ball, when

$$\langle m(\mathbf{P}) m^*(\mathbf{P'}) \rangle = \frac{1}{a} M(\mathbf{P}) \delta(\mathbf{P}-\mathbf{P'}), \tag{36}$$

Using the spherical coordinate, transverse coherence function for uniform sources density can be presented as

$$W(R,\alpha) = \frac{M}{(4\pi)^2 a} \int_0^a P^2 dP \int_{-\pi}^{\pi} d\varphi \int_{-\pi/2}^{\pi/2} \frac{\cos\theta \, d\theta \, \exp\{ik[p(\alpha)-p(-\alpha)]\}}{p(\alpha)p(-\alpha)}, \tag{37}$$

where we used a shorthand notation

$$p(\alpha) \equiv \left[ P^2 + R^2 - 2PR\left( \cos\varphi \cos\theta \sin\frac{\alpha}{2} + \sin\theta \cos\frac{\alpha}{2} \right) \right]^{1/2}. \tag{38}$$

Wave irradiance is calculated as

$$I(R) = W(R,0) = \frac{M}{4\pi^2 a} \int_0^a P^2 dP \int_{-\pi}^{\pi} d\varphi \int_{-\pi/2}^{\pi/2} \frac{\cos\theta \, d\theta}{p^2 + R^2 - 2pR\sin\theta\cos\frac{\alpha}{2}}$$

$$= \frac{M}{8\pi}\left[1 - \frac{(R^2 - a^2)}{2aR}\ln\left(\frac{R+a}{R-a}\right)\right].$$ (39)

For $R \gg a$

$$I(R \gg a) \approx \frac{Ma^2}{12\pi R^2},$$ (40)

and Eq. (37) by using expansions similar to Eq. (18) and the far-field DOC with the help of Eq. (6.567.13) of [14] simplifies to

$$w_{FF}(\alpha) = \frac{3}{2a^3}\int_0^a P^2 dP \int_{-1}^1 dt J_0\left(kPt\sin\frac{\alpha}{2}\right) = \frac{3 j_1(\gamma)}{\gamma} = \frac{3}{\gamma^2}\left(\frac{\sin\gamma}{\gamma} - \cos\gamma\right), \quad \gamma = 2ka\sin\frac{\alpha}{2}.$$ (41)

For $ka > 1$ $w(\alpha)$ maintains finite values only for $\alpha < 1$. In this case Eq. (41) can be simplified to Eq. (29) derived for the normal monopole on a sphere. Fig. 10 shows examples of DOCs for monopole sources in the ball calculated by numerical integration of Eq. (37) and far-field asymptotes, Eq. (41).

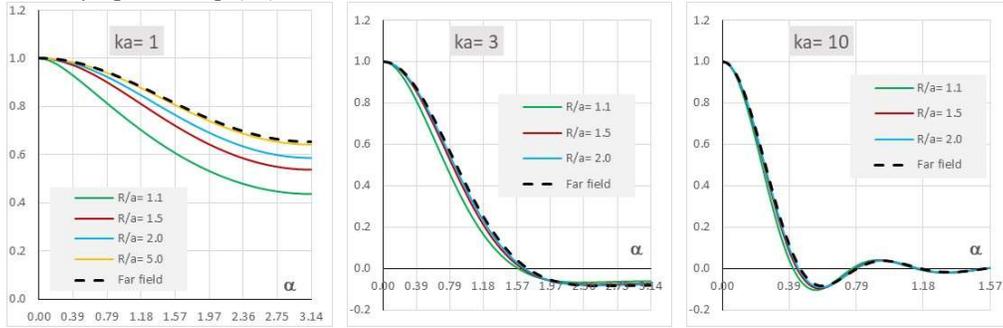

Fig. 10. Solid curves - numerically calculated DOC for monopole sources in a ball, Eq. (37), for three values of $ka$. Dashed curves – far-field result, Eq. (41).

Fig. 11 shows the examples of the far-field DOCs, Eq. (41) as solid curves and approximation, Eq. (29) for different values of $ka$. As expected, approximation is adequate for the central part of DOC when $ka > 1$.

### 3.2 Isotropic dipoles

Here we consider randomly oriented dipole sources inside a ball with radius $a$. Field at the point $\mathbf{R}$ outside the sphere can be presented as

$$U(\mathbf{R}) = \frac{1}{4\pi} \iiint_{P<a} d^3 P R_S \frac{d_I(\mathbf{P})\hat{\mathbf{d}}_I(\mathbf{P}) \cdot (\mathbf{P} - \mathbf{R})}{|\mathbf{R} - \mathbf{P}|^3}\left[1 - ik|\mathbf{R} - \mathbf{P}|\right]\exp(ik|\mathbf{R} - \mathbf{P}|).$$ (42)

Here $d_I(\mathbf{P})$ is random complex amplitude, and $\hat{\mathbf{d}}_I(\mathbf{P})$ is a random orientation of dipole momentum at the point $\mathbf{P}$ inside the ball. Assuming uniform distribution of incoherent complex amplitudes in the ball, similar to Eq. (36), and independent isotropic distribution of orientations, we calculate coherence function as

$$W(\mathbf{R}_1,\mathbf{R}_2)=\frac{D_I a}{48\pi^2}\int_0^a P^2 dP \int_{-\pi}^{\pi} d\varphi \int_{-\pi/2}^{\pi/2} \frac{d\theta\cos\theta(\mathbf{R}_1-\mathbf{P})\cdot(\mathbf{R}_2-\mathbf{P})}{|\mathbf{R}_1-\mathbf{P}|^3|\mathbf{R}_2-\mathbf{P}|^3}$$
$$(1-ik|\mathbf{R}_1-\mathbf{P}|)(1+ik|\mathbf{R}_2-\mathbf{P}|)\exp(ik|\mathbf{R}_1-\mathbf{P}|-ik|\mathbf{R}_2-\mathbf{P}|).$$
(43)

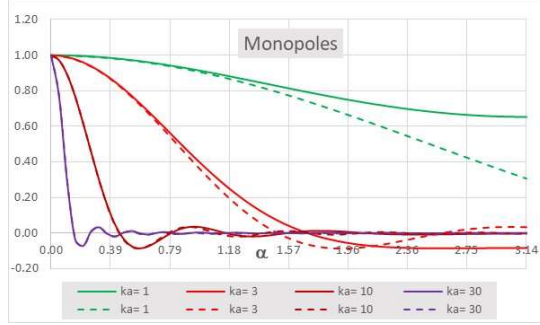

Fig. 11. Far-field DOC $w_{FF}(\alpha)$ for monopole sources in a ball for different values of $ka$. Solid curves, Eq. (41), dashed lines paraxial approximation, Eq. (29).

Irradiance can be calculated from Eq. (43) for $\mathbf{R}_1=\mathbf{R}_2$ as

$$I(R)=W(\mathbf{R},\mathbf{R})=\frac{D_I a^4}{24\pi}\int_0^a P^2 dP \int_{-1}^{1} dt \left[\frac{1}{(a^2+R^2-2aRt)^2}+\frac{k^2}{(a^2+R^2-2aRt)}\right]$$
$$=\frac{D_I a}{24\pi r}\left[\left(\frac{aR}{R^2-a^2}+\frac{1}{2}\ln\left(\frac{R+a}{R-a}\right)\right)+k^2 r^2\left(\frac{a}{R}-\frac{R^2-a^2}{R^2}\ln\left(\frac{R+a}{R-a}\right)\right)\right].$$
(44)

For $R\gg a$ Eq. (44) reduces to

$$I(R\gg a)\approx \frac{D_I}{36\pi}\frac{a^4}{R^4}(1+k^2 R^2).$$
(45)

Transverse DOC for isotropic dipoles when can be simplified for $R\gg a$ by using expansions, Eq. (18) and calculated analytically as

$$w_{FF}(\alpha)=3\cos\alpha\frac{j_1(\gamma)}{\gamma}=3\cos\alpha\left(\frac{\sin\gamma}{\gamma^3}-\frac{\cos\gamma}{\gamma^2}\right), \quad \gamma=2ka\sin\frac{\alpha}{2}$$
(46)

For $ka>1$ $w(\alpha)$ maintains finite values only for $\alpha<1$. In this case Eq. (46) can be simplified to the paraxial normal dipoles and ball monopole sources result, Eq. (29).

Figure 12 shows examples of DOCs for isotropic dipoles in a ball calculated by numerical integration of Eq. (43) for three values of parameter $ka$ as solid curves and the far-field asymptotes, Eq. (46), as dashed curves.

Fig. 13 shows the examples of the wide-angle far-field DOC, Eq. (46) and the paraxial approximation, Eq. (29), for different values of $ka$. As expected, Eq. (29) is adequate for the central part of DOC when $ka>1$.

Fig. 14 compares the far-field DOCs for the two model ball sources. Just as in the surface sources case, there are substantial difference between the models for moderate $ka$ values, but models become very similar with for large $ka$ values when $ka\alpha\propto 1$.

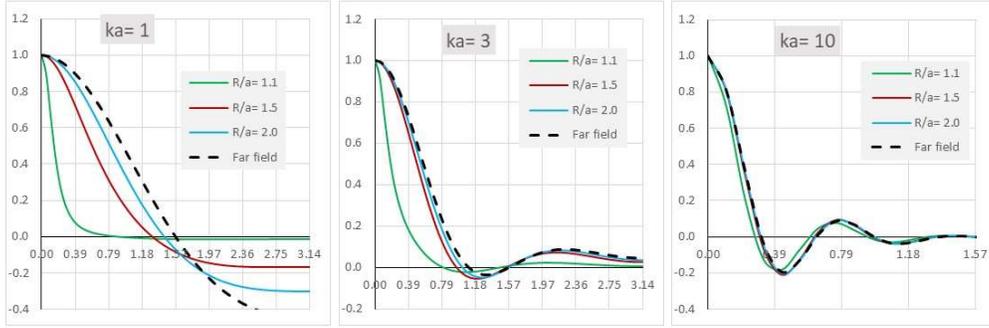

Fig. 12. Solid curves - numerically calculated DOC for incoherent isotropic dipoles on a sphere,
Eq. (43), for three values of $ka$. Dashed curves – far-field result, Eq. (29).

## 4. Plane disk

In this section we analyze coherence of the field created by incoherent sources distributed on a disc. This source type is commonly used in the literature to model coherence of solar light [1, 2, 3, 5, and 7]. Intuitively, it is less realistic than the spherical and ball models discussed in the previous section, but it is instructive to compare these three classes of models.

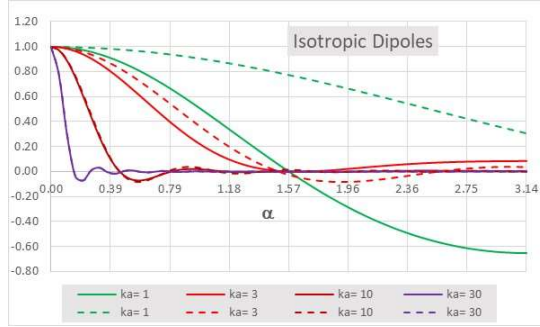

Fig. 13. Far-field DOC $w_{FF}(\alpha)$ for incoherent isotropic dipoles in a ball for different values of $ka$.
Solid curves, Eq. (46), dashed lines paraxial approximation, Eq. (29).

### 4.1 Monopole sources

Here we consider monopole sources located at a disk with radius $a$ at the plane $z=0$. Field at the point $\mathbf{R}$ outside the disk can be presented as

$$U(\mathbf{R}) = \frac{-1}{4\pi} \iint_{|R_S| \leq a} d^2 R_S \frac{m(\mathbf{R}_S)}{|\mathbf{R}-\mathbf{R}_S|} \exp(ik|\mathbf{R}-\mathbf{R}_S|), \quad \textbf{(47)}$$

where $m(\mathbf{R}_S)$ is complex amplitude of the source at the point $\mathbf{R}_S$. It is easy to show that $U(\mathbf{R}) = U(x,y,z)$ is an even function of $z$, and surface values of the field are non-locally related to the sources distribution as

$$U(x,y,0) = \frac{-1}{4\pi} \iint_{x_S^2+y_S^2 \leq a^2} \frac{dx_S dy_S m(x_S,y_S)}{\left[(x-x_S)^2+(y-y_S)^2\right]^{1/2}} \exp\left(ik\left[(x-x_S)^2+(y-y_S)^2\right]^{1/2}\right). \quad \textbf{(48)}$$

Note that this double integral has integrable singularity at $(x_S, y_S)$. In contrast, normal derivative of the field is an odd discontinuous function of $z$, and is locally related to the monopoles distribution

$$\left.\frac{\partial U(x,y,z)}{\partial z}\right|_{z\to 0} \approx \frac{z}{4\pi} \iint_{x_S^2+y_S^2\le a^2} \frac{dx_S dy_S m(x_S,y_S)}{\left[(x-x_S)^2+(y-y_S)^2+z^2\right]^{3/2}}\bigg|_{z\to 0} = \frac{m(x,y)}{2}\operatorname{sign}(z). \quad (49)$$

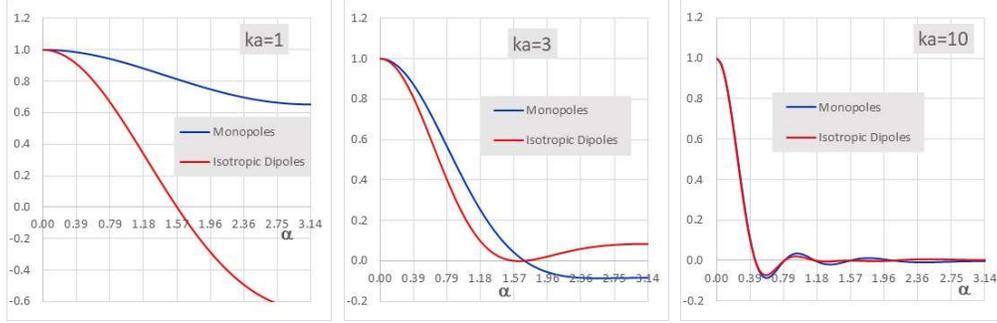

Fig. 14. Far-field DOC for three types of incoherent sources in a ball, Eq. (41), Eq. (47) and Eq. (46) for different values of $ka$.

This suggests that the incoherent monopole sources distribution would not create a delta-correlated field in the source plane. For the uniform distribution of incoherent sources

$$\langle m(\mathbf{R}_S)m^*(\mathbf{R'}_S)\rangle = M\delta(\mathbf{R}_S - \mathbf{R'}_S), \quad (50)$$

the two-point coherence function of the field is presented, as an integral over the circle

$$W(\mathbf{R}_1,\mathbf{R}_2) = \frac{M}{16\pi^2}\int_0^a r\,dr\int_0^{2\pi}\frac{d\theta}{|\mathbf{R}_1-\mathbf{R}_S||\mathbf{R}_2-\mathbf{R}_S|}\exp(ik|\mathbf{R}_1-\mathbf{R}_S| - ik|\mathbf{R}_2-\mathbf{R}_S|), \quad (51)$$

In general, this coherence function is neither isotropic, nor homogeneous. Here, keeping in mind application to the solar light coherence, we are only interested in the case when points $\mathbf{R}_{1,2}$ are symmetrical relative to the axis $z=0$, and can be prescribed by Eq. (15). In this case $W(\mathbf{R}_1,\mathbf{R}_2)=W(R,\alpha)$, and axial wave irradiance is

$$I(R) = W(R,0) = \frac{M}{16\pi}\ln\left(1+\frac{a^2}{R^2}\right)_{R\gg a} \approx \frac{M}{16\pi}\frac{a^2}{R^2}. \quad (52)$$

And axial transverse DOC is

$$w(R,\alpha) = \frac{W(R,R,\alpha)}{I(R)} = \frac{1}{\pi\ln\left(1+\dfrac{a^2}{R^2}\right)}\int_0^{2\pi}d\varphi\int_0^a\frac{r\,dr}{q(\alpha)q(-\alpha)}\exp[ikq(\alpha)-ikq(-\alpha)], \quad (53)$$

where we used a shorthand notation

$$q(\alpha) \equiv \left[r^2 + R^2 - 2rR\cos\varphi\sin\frac{\alpha}{2}\right]^{1/2}. \quad (54)$$

Fig. 15 presents examples of DOC calculated by numeric integration of Eq. (53). For $R\gg a$ Eq. (53) can be simplified by using expansions

$$q(\alpha)q(-\alpha) = R^2\left[1+O\left(\frac{a}{R}\right)\right], \quad q(\alpha)-q(-\alpha) = -2\rho\cos\varphi\sin\frac{\alpha}{2}\left[1+O\left(\frac{a^2}{R^2}\right)\right], \quad (55)$$

and DOC, Eq. (53) simplifies to

$$w_{FF}(\alpha) = \frac{1}{\gamma} J_1(\gamma), \quad \gamma \equiv 2ka\sin\frac{\alpha}{2}, \tag{56}$$

which is just a classic DOC of [1]. Dashed curves in Fig. 15 show far-field results, Eq. (56). For $ka > 1$ DOC maintains finite values only for $\alpha < 1$. In this case Eq. (56) can be simplified to the paraxial form

$$w_{PA}(\alpha) = \frac{2J_1(ka\,\alpha)}{ka\,\alpha}. \tag{57}$$

Fig. 16 shows the examples of the wide-angle far-field DOC, Eq. (56) as solid curves and paraxial approximation, Eq. (57) for different values of $ka$. As expected, paraxial approximation is adequate for the central part of DOC when $ka > 1$.

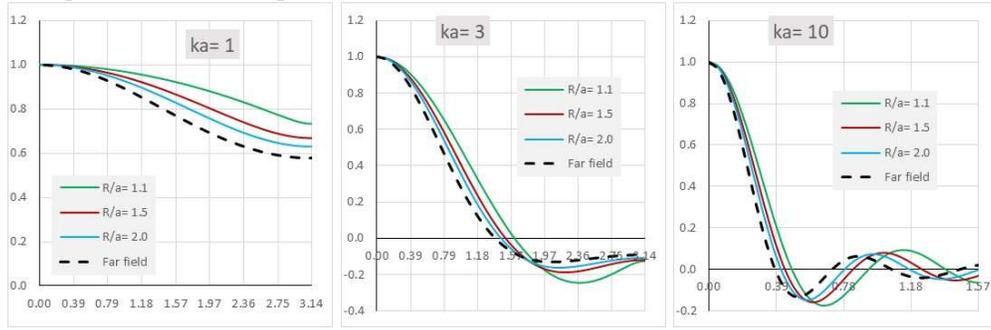

Fig. 15. Solid curves - numerically calculated DOC for monopole sources on a disk, Eq. (53), for three values of $ka$. Dashed curves – far-field result, Eq. (56).

## 4.2 Normal dipole sources

Here we consider dipole sources located at the circle with radius $a$. Dipole axis is assumed to be normal to the surface. Field at the point $\mathbf{R}$ outside the sphere can be presented as

$$U(\mathbf{R}) = \frac{1}{4\pi} \iint_{|R_S| \leq a} d^2 R_S \frac{d_N(\mathbf{R}_S)(\mathbf{R}_S - \mathbf{R}) \cdot \hat{\mathbf{z}}}{|\mathbf{R} - \mathbf{R}_S|^3} \exp(ik|\mathbf{R} - \mathbf{R}_S|)[1 - ik|\mathbf{R} - \mathbf{R}_S|], \tag{58}$$

where $d_N(\mathbf{R}_S)$ is complex amplitude of dipole at the point $\mathbf{R}_S$. In this case the surface field is locally related to the complex amplitude of dipole sources, namely

$$U(x,y,z)\big|_{z \to 0} \approx \frac{z}{4\pi} \iint_{|R_S| \leq a} dx_S dy_S \frac{d_N(x_S, y_S)}{\left[(x-x_S)^2 + (y-y_S)^2 + z^2\right]^{3/2}}\bigg|_{z \to 0} = \frac{-d_N(x,y)}{2} sign(z). \tag{59}$$

Hence delta-correlated normal dipole sources create delta-correlated fields at both sides of the circle, and this model should match the widely used [2, 3, 7] incoherent field model. Assuming uniform distribution of incoherent sources on the surface,

$$\langle d_N(\mathbf{R}_S) d_N^*(\mathbf{R'}_S) \rangle = a^2 D_N \delta(\mathbf{R}_S - \mathbf{R'}_S), \tag{60}$$

the two-point coherence function of the field is presented as an integral over the circle

$$W(\mathbf{R}_1, \mathbf{R}_2) = \frac{a^2 D_N R_1 \cos\alpha_1 R_2 \cos\alpha_2}{16\pi^2}$$
$$\times \int_0^a r\,dr \int_0^{2\pi} \frac{d\theta [1-ik|\mathbf{R}_1-\mathbf{R}_S|][1+ik|\mathbf{R}_2-\mathbf{R}_S|]}{|\mathbf{R}_1-\mathbf{R}_S|^3 |\mathbf{R}_2-\mathbf{R}_S|^3} \exp(ik|\mathbf{R}_1-\mathbf{R}_S|-ik|\mathbf{R}_2-\mathbf{R}_S|). \quad (61)$$

This coherence function is neither isotropic, nor homogeneous. As in the monopole case, we consider symmetric relative to the axis $z=0$ points, when $W(\mathbf{R}_1,\mathbf{R}_2)=W(R,\alpha)$.

Axial wave irradiance is

$$I(R) = W(R,0) = \frac{D_N a^4}{16\pi}\left(\frac{a^2+2R^2}{2R^2(a^2+R^2)^2} + \frac{k^2}{(a^2+R^2)}\right)_{R\gg a} \approx \frac{D_N a^4}{16\pi R^4}(1+k^2 R^2), \quad (62)$$

and similar to the spherical dipole sources irradiance consists of static and "wave" components. Axial transverse DOC is

$$w(R,\alpha) = \frac{2R^4(a^2+R^2)^4 \cos^2\frac{\alpha}{2}}{\pi a^2[a^2+2R^2+2k^2R^2(a^2+R^2)]}$$
$$\times \int_0^{2\pi} d\varphi \int_0^a \frac{r\,dr[1-ikq(\alpha)][1+ikq(-\alpha)]}{q^3(\alpha)q^3(-\alpha)} \exp[ikq(\alpha)-ikq(-\alpha)], \quad (63)$$

where $q(\alpha)$ is given by Eq. (54). Fig. 17 presents examples of DOC calculated by numeric integration of Eq. (63). For $R \gg a$ Eq. (63) can be simplified by using expansions, Eq. (55) as

$$w(\alpha) = \cos^2\frac{\alpha}{2}\frac{J_1(\gamma)}{\gamma}, \quad \gamma \equiv 2ka\sin\frac{\alpha}{2}. \quad (64)$$

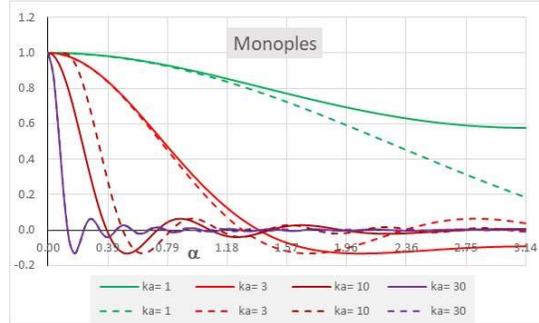

Fig. 16 Far-field DOC for incoherent monopoles on a disk $w(\alpha)$ for different values of $ka$.
Solid curves, Eq. (56), dashed lines paraxial approximation, Eq. (57).

Dashed curves in Fig. 17 show far-field results, Eq. (64). For $ka > 1$ DOC $w(\alpha)$ maintains finite values only for $\alpha < 1$, and paraxial approximation for normal dipoles is the same as for monopole sources, Eq. (57). Fig. 18 shows the examples of the wide-angle far-field DOCs, Eq. (64) as solid curves and paraxial approximation, Eq. (57) for different values of $ka$. As expected, paraxial approximation is adequate for the central part of DOC when $ka > 1$.

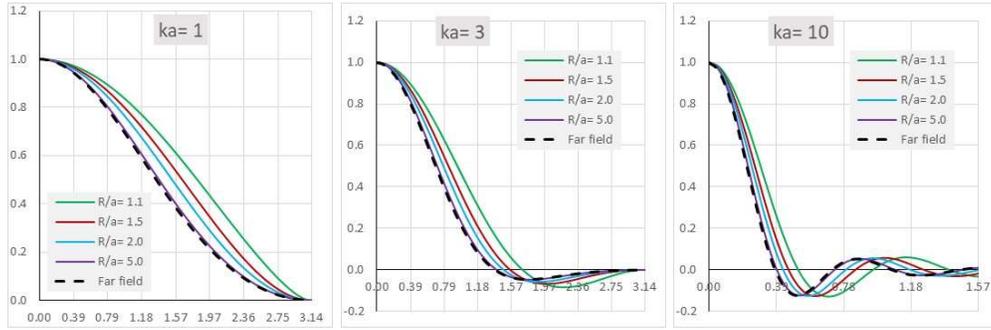

Fig. 17. Solid curves - numerically calculated DOC for normal dipole sources at a disk, Eq. (63), for three values of $ka$. Dashed curves – far-field result, Eq. (64).

### 4.3 Isotropic dipole sources

Here we consider randomly oriented dipole sources on a circle. Field at the point $\mathbf{R}$ outside the sphere can be presented as

$$U(\mathbf{R}) = \frac{1}{4\pi} \iint_{R_S < a} d^2 R_S \frac{d_I(\mathbf{R}_S) \hat{\mathbf{d}}_I(\mathbf{R}_S) \cdot (\mathbf{R}_S - \mathbf{R})}{|\mathbf{R} - \mathbf{R}_S|^3} \left[1 - ik|\mathbf{R} - \mathbf{R}_S|\right] \exp(ik|\mathbf{R} - \mathbf{R}_S|). \quad (65)$$

Here $d_I(\mathbf{R}_S)$ is random complex amplitude, and $\hat{\mathbf{d}}_I(\mathbf{R}_S)$ is a random orientation of dipole momentum at the point $\mathbf{R}_S$. In this case the surface field is locally related to the normally oriented dipole sources, namely

$$U(x, y, z)\big|_{z \to 0} \approx \frac{-1}{2} d_I(x, y) \sin\theta_d(x, y) \, \mathrm{sign}(z), \quad (66)$$

where $\theta_d(x, y)$ is the local elevation angle for dipole momentum. Again, delta-correlated randomly oriented dipole sources create delta-correlated fields at both sides of the circle, and this model should match incoherent field model [2, 3, and 7].

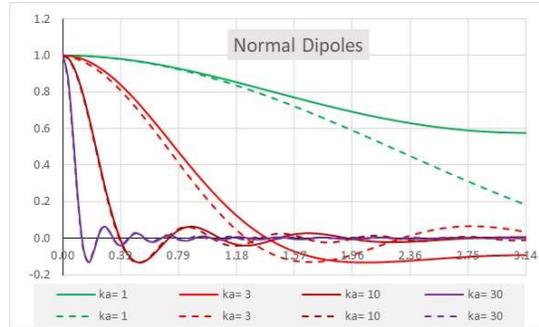

Fig. 18. Far-field DOC for normal dipoles on a disk $w(\alpha)$ for different values of $ka$. Solid curves, Eq. (64), dashed lines paraxial approximation, Eq. (57).

Assuming uniform distribution of incoherent complex amplitudes on the surface, similar to Eq. (60), and independent isotropic distribution of orientations, we calculate coherence function as

$$W(\mathbf{R}_1,\mathbf{R}_2) = \frac{a^2 D_N}{48\pi^2} \int_0^a r dr \int_0^{2\pi} \frac{d\theta (\mathbf{R}_1 - \mathbf{R}_S) \cdot (\mathbf{R}_2 - \mathbf{R}_S)}{|\mathbf{R}_1 - \mathbf{R}_S|^3 |\mathbf{R}_2 - \mathbf{R}_S|^3} \times (1 - ik|\mathbf{R}_1 - \mathbf{R}_S|)(1 + ik|\mathbf{R}_2 - \mathbf{R}_S|) \exp(ik|\mathbf{R}_1 - \mathbf{R}_S| - ik|\mathbf{R}_2 - \mathbf{R}_S|),$$ (67)

This coherence function is neither isotropic, nor homogeneous. Similar to the previous cases, further we consider symmetric relative to the axis $z=0$ points, when $W(\mathbf{R}_1,\mathbf{R}_2) = W(R,\alpha)$.

Axial wave irradiance is

$$I(R) = W(R,0) = \frac{D_I}{48\pi}\left(\frac{a^4}{R^2(a^2+R^2)} + k^2 a^2 \ln\frac{a^2+R^2}{R^2}\right)_{R\gg a} \approx \frac{D_I a^4}{48\pi R^4}(1+k^2 R^2),$$ (68)

and similar to the spherical dipole sources irradiance consists of static and "wave" components. Axial transverse DOC is

$$w(R,\alpha) = \frac{R^2(a^2+R^2)}{\pi\left(a^2 + k^2 R^2 (a^2+R^2)\ln\frac{a^2+R^2}{R^2}\right)} \int_0^{2\pi} d\varphi \int_0^a \frac{r dr (R^2 \cos\alpha + \rho^2)}{q^3(\alpha) q^3(-\alpha)}$$
$$\times [1 - ikq(\alpha)][1 + ikq(-\alpha)] \exp[ikq(\alpha) - ikq(-\alpha)],$$ (69)

where $q(\alpha)$ is given by Eq. (55). Fig. 19 presents examples of DOC calculated by numeric integration of Eq. (69). For $R \gg a$ Eq. (69) can be simplified by using expansions, Eq. (55) as

$$w(\alpha) = \cos\alpha \frac{J_1(\gamma)}{\gamma}, \quad \gamma \equiv 2ka\sin\frac{\alpha}{2}.$$ (70)

Dashed curves in Fig. 19 show far-field results, Eq. (70). For $ka>1$ $w(\alpha)$ maintains finite values only for $\alpha<1$, and paraxial approximation for normal dipoles is the same as for monopole DOC, Eq. (57). Wide angle far field results, Eq. (70) are shown in Fig. 20 as solid curves and paraxial approximation, Eq. (57) for different values of $ka$. As expected, paraxial approximation is adequate for the central part of DOC when $ka>1$.

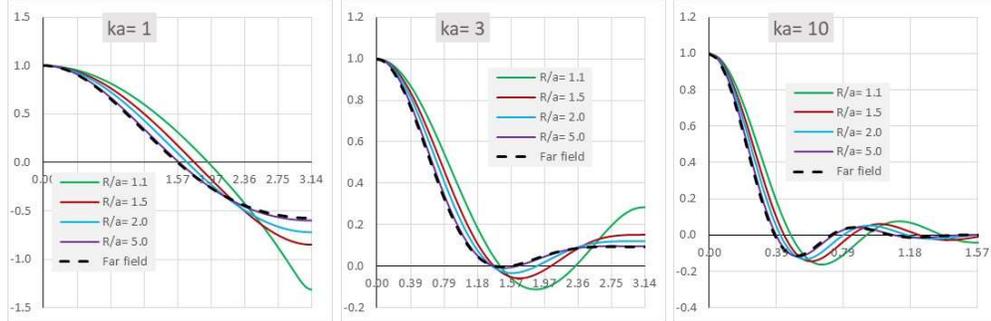

Fig. 19. Solid curves - numerically calculated DOC for isotropic dipole sources at a circle, Eq. (69), for three values of $ka$. Dashed curves – far-field result, Eq. (70).

## 5. Far field for incoherent sources

Here we examine the validity domain of the far-field approximation based on the case of monopoles on a sphere. Formally, as follows from Eq. (18), the asymptote represented by Eq. (19), is valid under conditions $R>a$ and $R>ka^2$. The later condition is the same as the far-field condition for a coherent source of size $a$. However, these constraints are not supported by calculation results. In Fig. 1 the dashed lines represent the far-field asymptotes, and, indeed, numerical results become closer to asymptote Eq. (19) when distance from the source increases for all source sizes. In the most interesting case, $ka > 1$, Fig. 1 suggests that the far-field Eq. (19) is valid close to the surface in contrast to the formal condition $R > ka^2$, suggested by derivation of Eq. (19). Detailed analysis of the difference between the exact DOC, Eq. (16) and the far-field asymptote for $R>a$ presented here reveals that the wide-angle far-field Eq. (19) is valid for $R>a$ uniformly in $\alpha$ and $ka$.

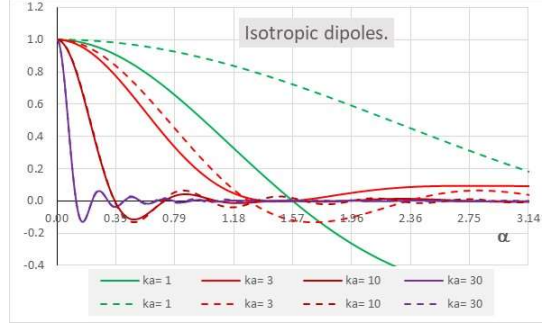

Fig. 20. Far-field DOC for incoherent monopoles on a circle $w(\alpha)$ for different values of $ka$.
Solid curves, Eq. (70), dashed lines paraxial approximation, Eq. (57).

Substitution of expansions, Eq. (18) in Eq. (16) allows to present the difference between the exact and far-field solution as

$$\Delta w(R,\alpha) \equiv w(R,\alpha) - w_{FF}(\alpha) = \frac{a^2}{R^2}\left[c_1(\alpha) + ka\sin\frac{\alpha}{2}c_2(\alpha) + k^2a^2\sin^2\frac{\alpha}{2}c_3(\alpha)\right] + O\left(\frac{a^4}{r^4}\right), \quad (71)$$

where

$$c_1(\alpha) = \frac{-1}{\pi}\int_0^\pi d\varphi \int_0^{\pi/2} \cos\theta d\theta \exp\left(-2ika\sin\frac{\alpha}{2}\cos\varphi\cos\theta\right) \\ \times \left(\frac{4}{3} - 2\sin^2\frac{\alpha}{2}\cos^2\varphi\cos^2\theta - 4\cos^2\frac{\alpha}{2}\sin^2\theta\right), \quad (72)$$

$$c_2(\alpha) = \frac{i}{\pi}\int_0^\pi d\varphi \int_0^{\pi/2} \cos\theta d\theta \exp\left(-2ika\sin\frac{\alpha}{2}\cos\varphi\cos\theta\right) \\ \times \cos\varphi\cos\theta\left(1 - 7\cos^2\frac{\alpha}{2}\sin^2\theta - \sin^2\frac{\alpha}{2}\cos^2\varphi\cos^2\theta\right), \quad (73)$$

and

$$c_3(\alpha) = \frac{-2}{\pi}\cos^2\frac{\alpha}{2}\int_0^\pi d\varphi \int_0^{\pi/2} \cos\theta d\theta \exp\left(-2ika\sin\frac{\alpha}{2}\cos\varphi\cos\theta\right)\cos^2\theta\sin^2\theta. \quad (74)$$

Integral expressions for coefficients $c_{1,2,3}(\alpha)$ can be further simplified, but it is irrelevant for the present discussion. We only note that $c_{1,2,3}(\alpha)$ are finite, $c_1(0)=c_2(0)=0$, and $c_3(0)=-2/15$.

For $ka \propto 1$ the bracketed factor in Eq. (71) is $O(1)$, which implies that the far-field is formed when $R > a$. Calculation results in Fig. 1 confirm this conclusion.

### 5.1. Central part of DOC for $ka > 1$

Practically more interesting case, $ka > 1$, requires a more detailed analysis. As is clear from Fig. 1 and Eq. (16), for $ka > 1$ DOC is significant only for $\alpha \propto (ka)^{-1} \ll 1$. In this case Eq. (71) can be simplified as

$$\Delta w\left(R, \alpha \propto \frac{1}{ka}\right) \approx \frac{a^2}{R^2}\left[\frac{1}{6}\frac{\sin(ka\alpha)}{(ka\alpha)} - \frac{1}{2}\frac{\sin(ka\alpha)}{(ka\alpha)^3} + \frac{1}{2}\frac{\cos(ka\alpha)}{(ka\alpha)^2}\right] \tag{75}$$

the bracketed factor in Eq. (75) is $O(1)$, which leads to the conclusion that for the central part of 19 is formed at $r > a$.

Fig. 21 shows the normalized difference of exact and far-field DOCs $\Delta w(R,\alpha) R^2/a^2$. Observed saturation at $r > 5a$ specifies that $\Delta w(R,\alpha) \propto a^2/R^2$, as predicted by Eq. (75). The oscillating differences $\Delta w(R,\alpha)$ typically have the maximum values at the slope of the main lobe of DOC, and we use these maximum differences as an estimate of the global error caused by the far-field approximation.

Fig. 22 shows the maximum normalized differences $\Delta w(R,\alpha) R^2/a^2$ between the exact and far-field DOCs as functions of the normalized radius $R/a$. As was noted above $\Delta w(R,\alpha) \propto a^2/R^2$ for $r > 5a$, but also the $ka$ dependence saturates for $ka > 5$ at the 0.05 value. This matches Eq. (75) where the bracketed expression has a maximum magnitude of 0.051 for $ka\alpha \approx 3.4$.

Finally, based both on numerical and analytical estimates, we conclude that the far field approximation for the central part of the coherence function is valid for $r > 5a$ with 0.2% accuracy when $ka > 5$.

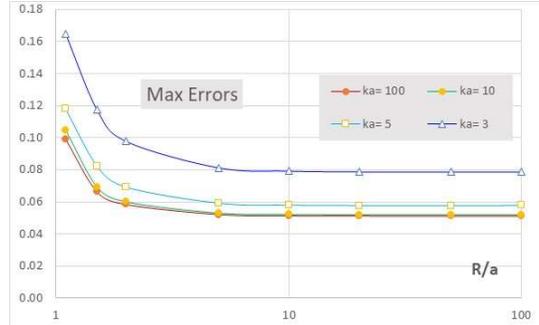

Fig. 22. Normalized maximal differences of exact and far-field DOC $\Delta w(R,\alpha) R^2/a^2$ for monopoles on a sphere for several values of $ka$.

### 5.2. Tail of the DOC for $ka > 1$

At the "tail" of DOC, $ka\sin(\alpha/2) > 1$, and $|w_{FF}(\alpha)| \propto 1/ka < 1$. Stationary phase method can be used to evaluate the integrals in Eqs. (72 - 74) leading to three critical points at the unit sphere

$$\varphi_1 = 0, \theta_1 = 0;\ \varphi_2 = \pi, \theta_2 = 0;\ \varphi_3 = \tfrac{\pi}{2}, \theta_3 = \tfrac{\pi}{2}. \tag{76}$$

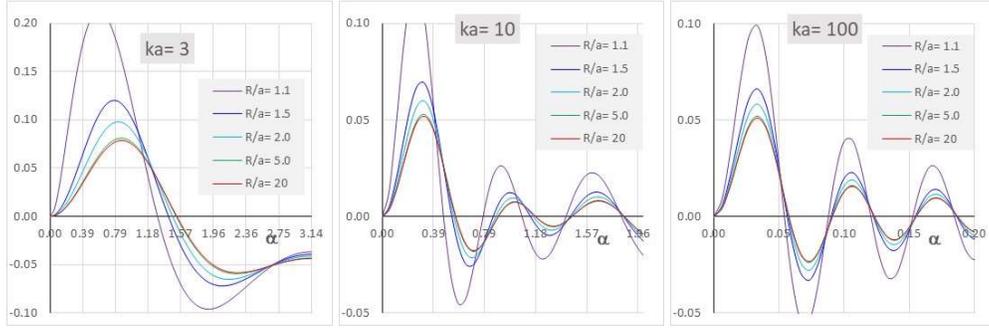

Fig. 21. Normalized difference of exact and far-field DOC $\Delta w(R,\alpha)R^2/a^2$ for monopoles on a sphere for three values of parameter $ka$ calculated by subtraction of the wide angle far-field result, Eq. (19) from numerically integrated Eq. (16).

Since here the effective integration area for the stationary phase method is $O(ka\sin(\alpha/2)) < 1$, at the first glance Eq. (71) suggests that

$$\Delta w(R,\alpha) = O\left(\frac{a^2}{R^2} ka \sin\frac{\alpha}{2}\right), \qquad (77)$$

leading to the far-field condition in the form $r > ka^2$. However, the detailed calculations show that the leading order terms of the asymptotic expansion of $c_2(\alpha)$ and $c_3(\alpha)$ at the first and second critical points cancel each other. This makes it necessary to account for the higher-order terms of the expansion in the vicinity of the first and second critical points. Turns out that the first and second critical points contribution to $c_1(\alpha)$ is $O[(ka\sin(\alpha/2))^{-1}]$, for $c_2(\alpha)$ it is $O[(ka\sin(\alpha/2))^{-2}]$, and for $c_3(\alpha)$ it is $O[(ka\sin(\alpha/2))^{-3}]$.

For the third critical point all three bracketed terms in Eq. (71) are $O[(ka\sin(\alpha/2))^{-3/2}]$, and its contribution is smaller than then from the first and second points. Finally, the difference between the exact and far-field DOC tails is

$$\Delta w(R,\alpha) = \frac{a^2}{R^2} w_{FF}(\alpha)\left(-\frac{1}{3} + \frac{1}{2}\cos^2\frac{\alpha}{2} + O\left(\left(ka\sin\frac{\alpha}{2}\right)^{-1/2}\right)\right) \qquad (78)$$

For the "tail" of the coherence function, specifically for $2\pi/ka < \alpha \leq 2\pi$, relative difference between the exact and far-field DOC is a more appropriate measure of the accuracy of the far-field approximation. Fig. 25 shows $(\Delta w(R,\alpha)/w_{FF}(\alpha))R^2/a^2$ as functions of separation angle $\alpha$ for $ka = 50$. Solid curves are based on the numerical integration of Eq. (71), and dashed curve is asymptotic result, Eq. (78). Fig. 23 confirms the validity of the conclusion that the far field is formed for $R > a$. Oscillations of the numeric data relative to the smooth analytic curve, Eq. (78) is related to the fact that, in contrast to the leading order term, Eq. (78), the higher-order term of $\Delta w(R,\alpha)$ are not proportional to $w_{FF}(\alpha)$, and small absolute values of $\Delta w(R,\alpha)R^2/a^2$ are enhanced in the vicinity of the $w_{FF}(\alpha)$ zeroes.

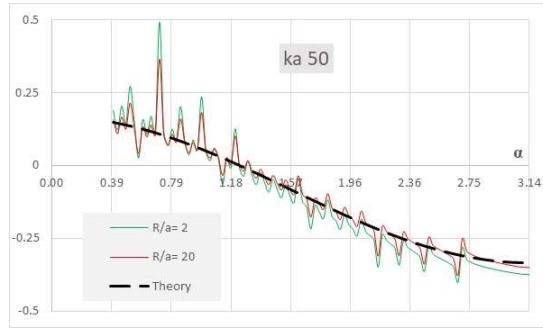

Fig. 23. Normalized relative difference between exact and far-field DOC for the "tail" of coherence function.

There are claims in the literature, [2, 12] that the far field for incoherent sources is formed at the distance of several wavelengths from the source. These claims are based on the visual impression given by the charts similar to Fig. 1, 4, 7, but not on any analytical or numerical examinations. This contradicts our conclusion that far field is formed at the distances larger than the source size. Fair resolution of this controversy requires estimations of $\Delta w(R,\alpha)$ for very large values of $ka$, since only in this case there exists a range of distances such that $k(R-a) \gg 1$, but still $R \propto a$. Series summation approach of [2] was limited to $ka \leq 100$, but direct numerical integration used here is capable to handle much larger $ka$ values. Fig. 24 shows $\Delta w(R,\alpha)$ for $ka = 3000$, and three source types on a sphere. At the minimum distance shown of about $50\lambda$, but $R = 1.1a$ the difference still can be as high as 20%, and only becomes small when it is exceeds the source size.

Finally, we conclude that the for an incoherent spherical source far-field coherence function is formed at the distances larger than the source radius independently of the wave length, and DOC has the form of Eq. (19). This conclusion also stands for the other source types considered here, as supported by the calculation data presented in Fig 1, 4, and 7. This is in contrast to the coherent sources case when far-field is formed when $R > ka^2$, or partially coherent case with finite coherence radius $r_C < a$, where the far-field condition is $R > kar_C$ [13]. In the far-field irradiance is proportional to the inverse square of distance, $I(R) \propto R^{-2}$, and transverse DOC of the field depends only on the angular separation of the observation points and is independent of the distance.

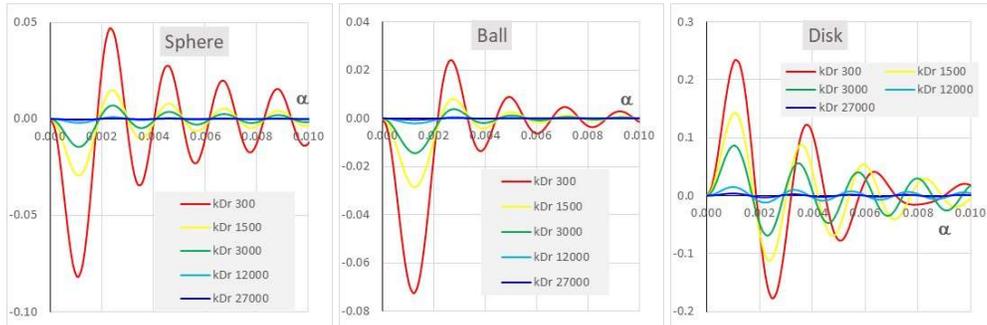

Fig. 24. Differences of exact and far-field DOC $\Delta w(R,\alpha)$ for monopoles on a sphere, in a ball and on a disk for $ka = 3000$. Parameter is $k(R-a)$.

## 6. Discussion

### 6.1 Three paraxial models

In the previous sections we used three sources configurations and three types of incoherent sources to introduce eight models for coherence of light waves. All models allow calculation of two-points coherence functions for any source sizes and arbitrary points positions by numeric integration of simple 2-D or 3-D integrals. It was found numerically, and confirmed analytically that the far-field for all types of sources considered is formed at the distances larger than the source size, $R > a$. For simple, single-scale source models considered here, a single dimensionless parameter for all far-field DOC is $ka$, and $w_{FF}(\alpha)$ have simple analytical forms, Eq. (19, 28, 34, 41, 47, 46, 56, 64, and 70). Figures 25 – 27 show all eight DOC for several $ka$ values.

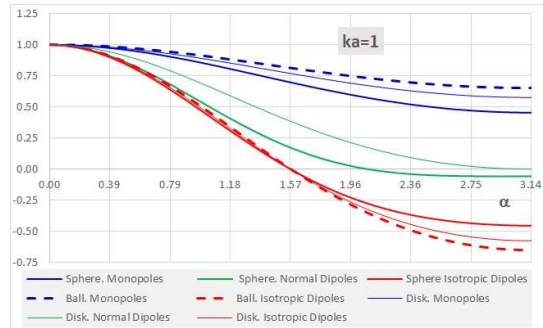

Fig. 25. Far-field DOC for the eight source models and $ka = 1$. Heavy solid curves, sources on a sphere Eq. (19, 28, 34), dashed lines sources in a ball, Eq. (41, 46).and thin lines – sources on a disk Eq. (56, 64, 70).

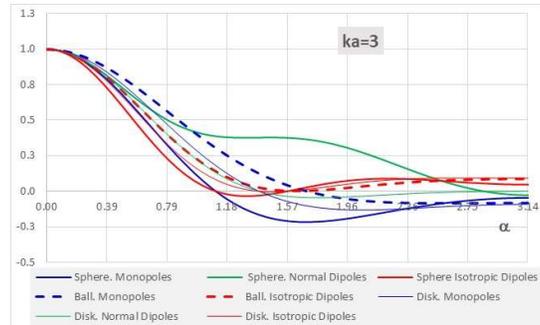

Fig. 26. Far-field DOC for the eight source models and $ka = 3$. Heavy solid curves, sources on a sphere Eq. (19, 28, 34), dashed lines sources in a ball, Eq. (41, 47, 46).and thin lines – sources on a disk Eq. (56, 64, 70).

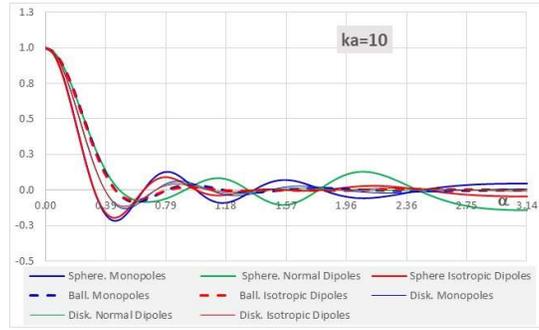

Fig. 27. Far-field DOC for the eight source models and $ka = 10$. Heavy solid curves, sources on a sphere Eq. (19, 28, 34), dashed lines sources in a ball, Eq. (41, 47, 46).and thin lines – sources on a disk Eq. (56, 64, 70).

It is clear, that for $ka \geq 10$ eight curves, at least for the central part of the DOC, merge into just three bundles, corresponding to
- monopoles and isotropic dipoles at a sphere,
- monopoles and isotropic dipoles in a ball and normal dipoles at a sphere,
- all types of sources on a disk,

These three main models can be represented by incoherent monopoles on a sphere, in a ball and on the disk, and are described by paraxial approximations, Eq. (20, 29, 63) that are valid for $R > 5a$ $ka \gg 1$, and $ka\,\alpha \propto 1$.

### Coherence of sunlight and limb darkening

Fig. 28 shows DOC for the solar light, $a_\circ = 6.96 \cdot 10^8\,m$, as would be measured on Earth at the distance $R_\circ = 1.50 \cdot 10^{11}\,m$. The first zero coherence length ranges from $107\lambda$ for the sphere to $154\lambda$ for the ball source. All disk models have a well-known zero at $131\lambda$ [1]. Notably, these data were obtained both by numerical integration of Eq. (16, 37, 53) and wide-angle far-field asymptotes, Eq. (19, 41, 56), proving that direct numerical integration is still efficient for $ka \propto 10^{16}$, as long $ka\alpha$ is not extremely large.

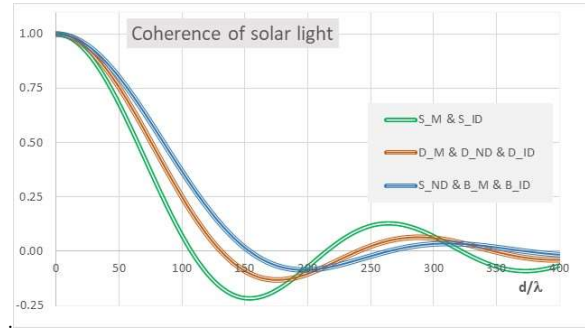

Fig. 28. DOC for the solar light as functions the linear point's separation. Green double curve: spherical source. Brown disk source. Triple blue curve: ball source.

Given the DOC of solar light, it is straightforward to calculate the image of the Sun that is formed by an optical system with aperture size much larger than coherence radius. Namely, image irradiance at the angular coordinate $\vec{\gamma}$ is Fourier transform of the far-field angular coherence

$$I_{IM}(\vec{\gamma}) = C \iint d^2\alpha\, W(\vec{\alpha}) \exp(ikR_\circ \vec{\alpha} \cdot \vec{\gamma}). \tag{79}$$

For the DOC given by Eq. (20), spherical source, the image is calculated using Eq. (6.671.7) of [14] as

$$I_{IM}(\vec{\gamma}) \propto \begin{cases} (1-\gamma^2/\gamma_\circ^2)^{-1/2}, & \gamma < \gamma_\circ \\ 0, & \gamma > \gamma_\circ \end{cases}, \qquad (80)$$

where $\gamma_\circ = a_\circ/R_\circ$ is the Sun angular radius. For the DOC given by Eq. (29) ball source the image is calculated using Eq. (6.693.1) of [14] as

$$I_{IM}(\vec{\gamma}) \propto \begin{cases} (1-\gamma^2/\gamma_\circ^2)^{1/2}, & \gamma < \gamma_\circ \\ 0, & \gamma > \gamma_\circ \end{cases}, \qquad (81)$$

and, finally for the disk model, Eq. (63) image is uniform, as expected

$$I_{IM}(\vec{\gamma}) = C \begin{cases} 1, & \gamma < \gamma_\circ \\ 0, & \gamma > \gamma_\circ \end{cases}, \qquad (82)$$

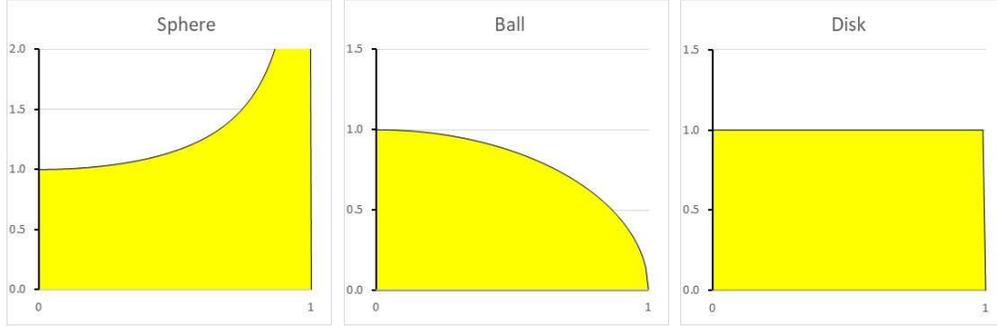

Fig. 29. Sun images cross-sections for three models Eqs. (20, 29, 63) as functions of $\gamma/\gamma_\circ$.

Fig. 29 shows $I_{IM}(\gamma/\gamma_\circ)$ irradiance profiles of Sun images corresponding to the sphere, ball and disk paraxial DOC models. Sphere model, Eq. (20) displays unphysical limb brightening, while the ball model presented by Eq. (29) shows some limb darkening, and has the same functional form as the first-order term of the series used in [8]. The limb darkening profile can be modified by using the ball sources with radial dependence of the sources density. For example if in Eq. (11)

$$M(\mathbf{P}) = M(1-P^2/a^2)^n, \qquad (83)$$

then corresponding image is

$$I_{IM}(\vec{\gamma}) \propto \begin{cases} (1-\gamma^2/\gamma_\circ^2)^{n+1/2}, & \gamma < \gamma_\circ \\ 0, & \gamma > \gamma_\circ \end{cases}, \qquad (84)$$

Linear combination of several terms given by eq. (84) results in the limb darkening function series similar to one used in [8] to match the solar radiation measurements.

## 7. Summary and conclusions

We investigated eight models for the DOC of radiation from incoherent spherical sources based in radiation of incoherent monopoles and dipoles distributions on a sphere, in a ball, and on a plane disk.

In contrast to many earlier works on the subject, [1 – 7] based on the incoherent boundary conditions and/or spherical harmonics series, we use Green's functions and superposition principle here. This leads to the double or triple integral DOC representation. These integrals can be easily calculated numerically for the wide ranges of the source sizes and observation point's positions. Multiple examples of these calculations are presented in the text.

We showed analytically and confirmed numerically that the far-field for the DOC is formed at the distances $r > 5a$ and is independent on the radiation wavelength. This is different from common coherent and partially coherent [13] estimates and in contradict the $r > 7\lambda$ estimate of [2].

In case of the large, $ka \gg 1$, sources paraxial approximation is sufficient in the far-field for the central part of the DOC, and eight DOC models reduce to just three, which can be represented by incoherent monopoles uniformly distributed on a sphere, in a ball, and on a disk. Corresponding coherent radii defined at the first DOC zero are $107\lambda$, $154\lambda$ and $131\lambda$.

Solar images corresponding to the spherical DOC shows unphysical limb brightening in contrast to the uniform image for the disk model, and more reasonable limb darkening for the ball DOC. This leads us to believe that the ball DOC, Eq. (29), is a more realistic model than the currently most common disk model, Eq. (63).

Ball model can be modified by introducing the radial dependence of the source amplitudes to match the observed profiles of the limb darkening, but this is outside of the scope of this work.

## References

1. M. Born and E. Wolf, *Principles of Optics*, 7th (expanded) ed. (Cambridge U. Press, Cambridge, 1999).
2. S. S. Agraval, G. Gbur, and E. Wolf, "Coherence properties of sunlight," Opt. Lett. **29**, 459-461 (2004)
3. R. Borghi, F. Gori, O. Korotkova, and M. Santarsiero, "Propagation of cross-spectral densities from spherical sources," Opt. Lett. **37**, 3183–3185 (2012).
4. F. Gori and O. Korotkova, "Modal expansion for spherical homogeneous sources," Optics Comm. **282**, 3859-3861 (2009)
5. H. Mashaal, and A. Goldstein, "Fundamental bounds for antenna harvesting of sunlight," Opt. Lett **36**, 900-902 (2011)
6. P. B. Lerner, P. H. Cutler, and N. M. Miskovsky, "Coherence properties of blackbody radiation and application to energy harvesting and imaging with nanoscale rectennas," J. Nanophotonics **9**, 093044, (2015).
7. S. Divitt and L. N, Novotny, "Spatial coherence of sunlight and its implications for light management in photovoltaics," Optica **2**, 95-103 (2015)
8. H. Neckel and D. Labs, "Solar limb darkening 1986–1990 (λλ303 to 1099 nm)," Sol. Phys. **153**, 91–114 (1994).
9. H. Mashaal, A. Goldstein, D. Feuermann, and J. M. Gordon, "First direct measurement of the spatial coherence of sunlight," Opt. Lett. **37**, 3516-3518 (2012)
10. J. Lindberg, T. Setälä, M. Kaivola, and A.T. Friberg, "Coherence and polarization properties of a three-dimensional, primary, quasi-homogeneous, and isotropic source and its far field," Opt Comm. **283**, 4452-4456, (2010).
11. J. R. Zurita-Sánchez, "Coherence properties of the electric field generated by an incoherent source of currents distributed on the surface of a sphere," J. Opt. Soc. Am. A **33**, 118-130 (2016).
12. S. Sundaram and P. K. Panigrahi, "On the origin of the coherence of sunlight on the earth," Opt. Lett. **41**, 4222–4224 (2016).
13. F. Gori, "Far-zone approximation for partially coherent Sources," Opt. Lett. **30**, 2840 (2005).
14. S.I. S. Gradshteyn and I. M. Ryzhik, *Table of Integrals, Series, and Products*. Seventh Edition, Elsevier, 2007